\documentclass[]{aastex7}
\usepackage{graphicx}
\graphicspath{{./}}

\begin{document}

\title{Ringed versus Ringless Worlds: How Poynting-Robertson Drag Shapes Rings across the Solar System}

\author[orcid=0000-0003-4590-0988]{Ryuki Hyodo}
\affiliation{Earth-Life Science Institute, Institute of Science Tokyo, 2-12-1 Ookayama, Meguro-ku, Tokyo, 152-8550, Japan}
\affiliation{Universit\'{e} Paris Cit\'{e}, Institut de Physique du Globe de Paris, CNRS, F-75005 Paris, France}
\affiliation{Graduate School of Artificial Intelligence and Science, Rikkyo University, Tokyo 171-8501, Japan}
\affiliation{SpaceData Inc.}
\email{ryuki.h0525@gmail.com}

\author[orcid=0000-0002-9676-3891]{Shigeru Ida}
\affiliation{Earth-Life Science Institute, Institute of Science Tokyo, 2-12-1 Ookayama, Meguro-ku, Tokyo, 152-8550, Japan}
\email{ida@elsi.jp}

\begin{abstract}
Planetary rings are not only ubiquitous around the giant planets in the outer Solar System, but have also been discovered around several small distant bodies. In contrast, no rings have been observed around any inner Solar System objects. To constrain the dynamical origin of this ringed-versus-ringless dichotomy, we employ a numerically cross-checked analytical model of gigayear-scale Poynting-Robertson (PR) drag due to the solar flux acting on an isolated particle, expressed as a function of the host body's heliocentric distance \(\,a_{\mathrm{pla}}\) and the particle radius \(\,r_{\mathrm{par}}\). Here we show that, in the absence of additional perturbations, PR drag alone can explain the observed ring architecture of the Solar System: outer planets and Centaurs/TNOs are able to retain rings for the age of the Solar System, whereas any rings around the inner planets are removed on much shorter timescales. Because the PR-drag lifetime scales steeply with heliocentric distance \(\bigl(\tau_{\mathrm{decay}}\propto a_{\mathrm{pla}}^{2} \,r_{\mathrm{par}}\bigr)\), we predict that forthcoming surveys will reveal an ever-growing population of ring-bearing bodies in the distant Solar System.
\end{abstract}



\section{Introduction}

Planetary rings are among the most striking --- but also the most selective --- features in the Solar System. Although a variety of mechanisms can generate rings, only a fraction of planetary bodies are actually observed to host them. As discussed below, a ringed-versus-ringless dichotomy appears to exist between the inner and outer Solar System. Understanding why rings persist around some objects but not others therefore remains a fundamental problem in planetary science.

One well-studied ring formation pathway around terrestrial planets in the inner Solar System is collisional. During the final stages of terrestrial planet formation, giant impacts inevitably eject debris that can be gravitationally captured by the surviving body \citep{Har95}. The canonical Earth-Moon-forming impact \citep[e.g.,][]{Can04,Kok00} and the putative giant impact that produced Phobos and Deimos \citep[e.g.,][]{Hyo17a,Ros16} have been studied in terms of the dynamical evolution of the resulting planetocentric debris: each would have left an optically thin particulate ring once satellite accretion was complete. Yet no such rings are seen around Earth or Mars today \citep{Sho06}. Similarly, Mercury and Venus lack both moons and rings --- perhaps because their impacts never produced disks/rings massive enough to form satellites, or because any satellites were later removed by tidal interactions \citep{Ato07}. Even smaller-scale impacts can in principle create transient debris rings, but none of the present-day terrestrial planets and small inner solar system bodies exhibit ring systems.

Various formation mechanisms have been proposed to explain the rings observed around bodies in the outer Solar System. For example, close encounters between a giant planet and a passing large body can tidally strip material from a passing body, seeding a ring around the planet \citep{Don91,Hyo17b}. Meanwhile, the remaining large body can gravitationally capture part of its own debris to acquire a ring of its own \citep{Hyo16}. This is just one example of ring formation, but such processes demonstrate that rings are not confined to massive planets in the outer Solar System. Indeed, at the time of this paper is written, particulate rings are confirmed or suspected around several small outer solar system bodies through observing stellar occultations --- including the Centaur Chariklo \citep{Bra14}, possibly Chiron \citep{Ort15}, the dwarf planet Haumea \citep{Ort17}, and the trans-Neptunian object Quaoar \citep{Mor23,Per23}. Statistical estimates suggest that rings occur around roughly $8-12$\% of small outer solar system bodies \citep{Sic20}, implying that additional ring-bearing objects will likely be discovered.

Indeed, the aforementioned ring-forming processes could operate in both the inner and outer Solar System; however, observations clearly indicate that rings are more prevalent in the outer Solar System. Whether produced by impacts, tidal encounters, or other processes, the ultimate survival of a ring depends on the dynamical evolution of its constituent particles. Accordingly, this evolution should be consistent with the observed ringed-versus-ringless dichotomy between the inner and outer Solar System.

The evolution of ring particles is governed by several distinct physical regimes. When a ring is optically thick, collisional and gravitational interactions among the particles can dominate, leading to viscous spreading \citep{Lyn72,Gol78}. As an alternative, when the central body's shape is non-axisymmetric, resonant effects arise between the object's spin rate and the ring particles' mean motion \citep{Sic20}.

Here we instead consider the optically thin limit and follow the dynamics of individual, isolated particles around a spherical central body due to non-gravitational forces. The dominant non-gravitational force on a ring particle depends on its size. Here we focus on "medium-sized" particles --- typically centimeter- to meter-sized --- which we assume are neither rapidly removed by solar radiation pressure, taking into account the planetary shadow \citep{Lia23}, nor strongly affected by Yarkovsky or YORP effects \citep{Rub00}. In this size regime, Poynting-Robertson (PR) drag due to the solar flux may dominate the non-gravitational evolution, gradually removing orbital energy and leading to a contraction of the semimajor axis \citep[see also][]{Bur01}. Note that if the planetary shadow is neglected, solar PR drag produces no secular change in eccentricity \citep{Bur79}; to the best of my knowledge, the giga-year scale long-term effect of the shadow on eccentricity has not yet been investigated.

Previous studies \citep[e.g.,][]{Bur79} carried out an extensive analytical study of Poynting-Robertson drag, but at that time rings had not yet been discovered around the wide range of body sizes and orbital radii we know today. Consequently, the theory could neither fully account for observational data nor be tightly constrained by it. In this present work, we revisit the problem in light of new observations to reassess the potential role of PR-drag in shaping the distribution of rings throughout the solar system.\\

\section{Methods} 
\label{sec_method}
In this section we describe the methodology of our numerical simulations. We concentrate exclusively on Poynting-Robertson (PR) drag due to the solar flux. To isolate the role of PR drag, we adopt a deliberately simplified approach. We assume that an isolated ring particle is present at around a spherical host body and then follow its dynamical evolution over gigayear timescales. Our results are presented as functions of particle size $r_{\rm par}$ and the radial distance of the central body from the Sun $a_{\rm pla}$, providing a baseline for assessing which rings can plausibly persist to the present day and which are destined to vanish on shorter timescales.

A direct, first-principles integration of the PR-drag force over gigayear timescales is computationally prohibitive, so we adopt instead the analytical prescription recently developed by \citet{Lia23}. Using classical planetary perturbation theory \citep{Mur99} together with the canonical formulation of PR drag \citep{Bur79}, \citet{Lia23} (their Eq.~(20)) derived an expression for the \textit{lowest} secular decay rate of a particle's semimajor axis about a spherical central body, whether a planet or a small object, on a circular orbit:
\begin{eqnarray}
\label{eq_LH23_Eq20}
	\left\langle \frac{da_{\rm par}}{dt} \right\rangle && = -\frac{2\,B |v_{\rm par}| }{n_{\rm par}\,c|d|^{2}} \Biggl[ 1 + \frac{1}{4}\!\left(1+\cos^{2} i\right)  - \frac{1}{2\pi}
		\!\left(
  			\frac12 \left(1+\cos^{2} \epsilon \right) \phi + 2\phi +  \frac12 \left( 5-\cos^{2} \epsilon \right) \sin 2 \phi
		\right)
	\Biggr] ,
\end{eqnarray}
\begin{equation}
	B \;=\; \frac{\pi\, Q\, r_{\mathrm{par}}^{2}\, F_{\mathrm{pla}}\, a_{\mathrm{pla}}^{2}} {c\, m_{\mathrm{par}}},
\label{eq_B}
\end{equation}
\begin{equation}
	F_{\mathrm{pla}}\!\left(a_{\mathrm{pla}}\right) \;=\; \frac{L_{\odot}}{4\pi\, a_{\mathrm{pla}}^{2}} ,
\label{eq_F}
\end{equation}
where $a_{\rm par}$ is the particle's semimajor axis; definitions of all other variables appearing in the expression are summarized in Table\,\ref{table_1}. This newly derived equation explicitly incorporates the angular extent of the planetary shadow, \(2\phi\) (see Fig.~1 in \citealt{Lia23}), as well as the effects of particle inclination \(i\) and planetary obliquity \(\epsilon\).

Here, \textit{lowest} signifies that Eq.~\ref{eq_LH23_Eq20} is derived under the assumption of a particle's circular orbit ($e = 0$). In reality, perturbations can excite a finite eccentricity, so that a particle's pericentre lies inside its semimajor axis; therefore, the resulting smaller pericentre distance makes it easier for the particle to collide with the planet. Direct numerical integrations that incorporate the planetary shadow, planetary oblateness \(J_2\), solar radiation pressure, and the solar Poynting-Robertson drag show that a particle's orbit becomes non-circular and the decay rate given by Eq.~\ref{eq_LH23_Eq20} is indeed a lower bound on the orbital evolution toward impact with the central body \citep{Lia23}. 

The actual lifetime before collision with the host planetary body depends on the attained eccentricity, because the pericentre distance is $q = a(1-e)$. For particles that start on circular orbits, larger objects are less susceptible to eccentricity growth driven by non-gravitational forces than smaller ones; consequently, their lifetimes more closely approach the lower bound estimate of Eq.~\ref{eq_LH23_Eq20}, and vice versa \citep{Lia23}.

In this study, using the above equation, we examine how planetary parameters --- specifically the planet's heliocentric distance and mass --- and particle size affect the PR-drag lifetime of circumplanetary particles. Quantifying these dependencies is the primary objective of this study, and we also discuss which bodies in the Solar System could host rings.

\section{Results: Ringed versus Ringless Worlds} 
\label{sec_results}

\begin{figure}[htbp]
	\begin{center}
	\includegraphics[width=0.5\textwidth]{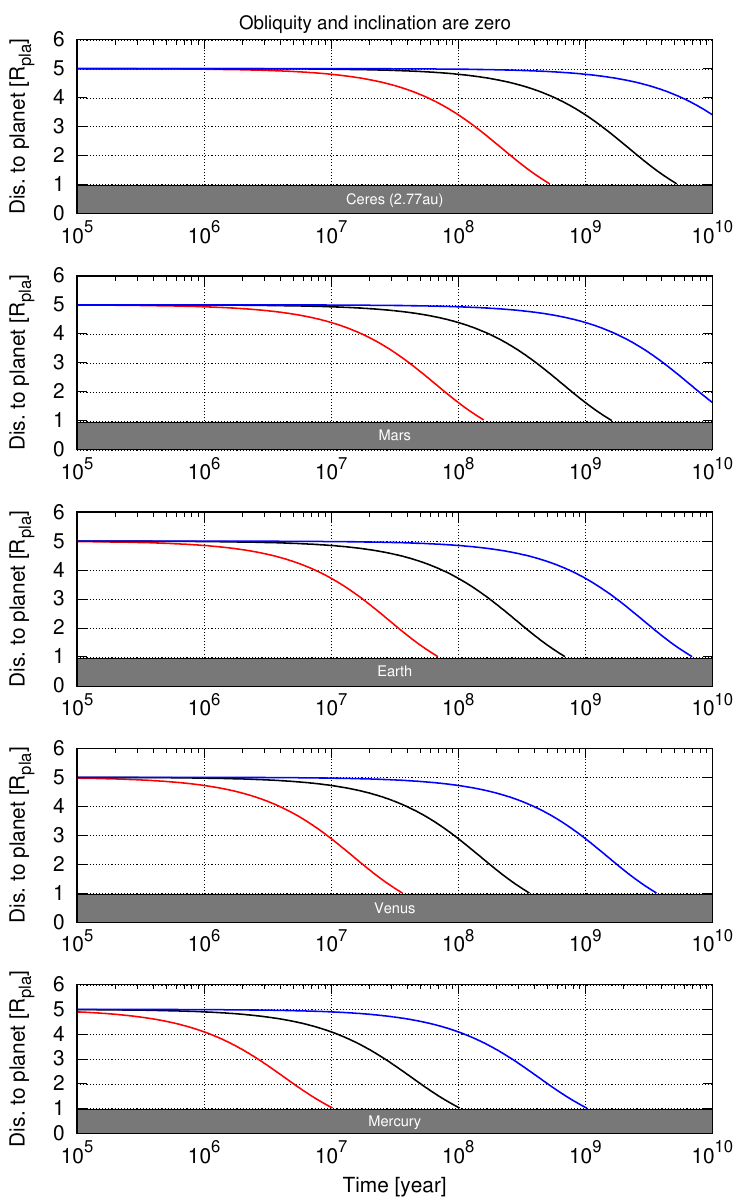}
	\caption{Time evolution of the orbital distance of a particle around a central body, initially starting from $a_{\rm par} = 5R_{\rm pla}$. For reference, $a_{\rm par} \sim 3R_{\rm pla}$ is the Roche limit of Earth and Mars. The red, black, and blue curves corresponds to a particle radius of $r_{\rm par}=1$\,cm, $10$\,cm, and $1$\,m, respectively. Both the planetary obliquity and the particle's orbital inclination are set to zero. In these calculations, the central body is assumed to be spherical. The calculations here include the perturbations from the planet's shadow (the $\phi$ term given in Table~\ref{table_1})}.
	\end{center}
\label{Fig_evolution}
\end{figure}

Figure~1 illustrates the time evolution of the particle's semimajor axis $a_{\rm par}$ for selected different host planetary bodies. Poynting-Robertson drag removes angular momentum from an orbit around a central body, causing the particle to spiral inward. As expected, particles that orbit host bodies located farther from the Sun, as well as particles with larger radii, survive for longer periods before ultimately colliding with the central body. In the figure, each particle is initially placed at $5R_{\mathrm{pla}}$, where $R_{\mathrm{pla}}$ denotes the radius of the central body. For comparison, the Roche limit for Earth and Mars is $\sim 3R_{\mathrm{pla}}$. Particles that begin at smaller orbital distances collide with the central body on correspondingly shorter timescales. 

\begin{figure*}[htbp]
\begin{center}
	\includegraphics[width=0.8\textwidth]{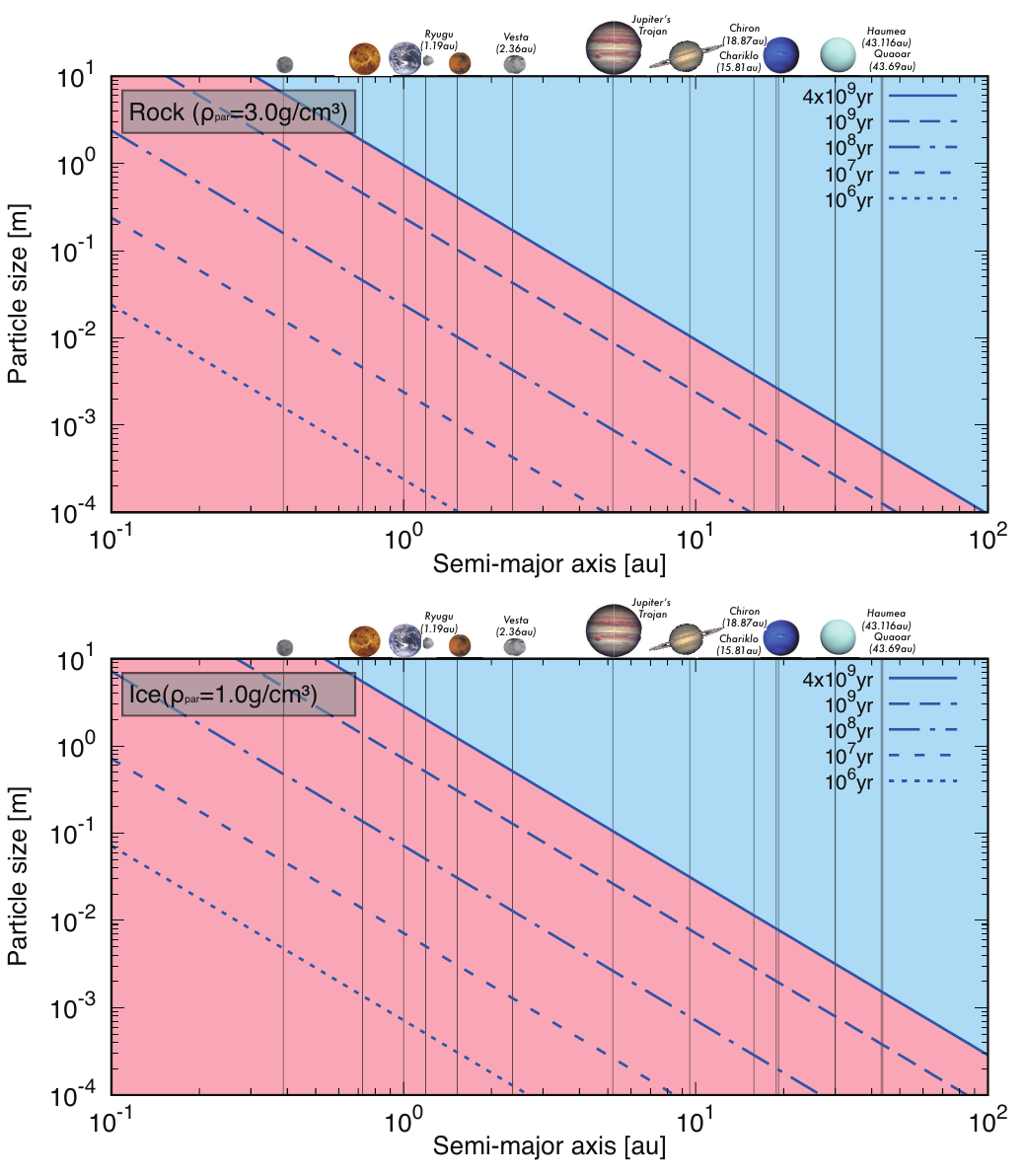}
\caption{Parameter space of the decay timescale as a function of the central body's semi-major axis and ring particle radius (Eq.~\ref{eq_decay}). The top and bottom panels correspond to cases with $\rho_{\rm par} = 3.0,{\rm g,cm^{-3}}$ and $\rho_{\rm par} = 1.0,{\rm g,cm^{-3}}$, corresponding to densities typical of rocky and icy ring particles, respectively. Regions with $\tau_{\mathrm{decay}} > 4\times10^{9}$\,yr are shaded light blue, whereas those with $\tau_{\mathrm{decay}} < 4\times10^{9}$\,yr are shaded light red. Contours indicate constant timescales of $\tau_{\mathrm{decay}} = 10^{9}$, $10^{8}$, $10^{7}$, and $10^{6}$\,yr. $Q=1$ is assumed. Vertical lines indicate the reference semi-major axes of selected planetary bodies.}
\label{fig_2}
\end{center}
\end{figure*}

The characteristic timescale for PR-drag-induced decay of a particle's semimajor axis \(\tau_{\mathrm{decay}}\) can be written as
\begin{eqnarray}
\label{eq_decay}
	&&\tau_{\rm decay} = 
	a_{\rm par} / \left( \frac{da_{\rm par}}{dt} \right) = 
	\frac{8 \pi c^2 \rho_{\rm par} r_{\rm par} a_{\rm pla}^2}{3 Q L_{\odot}} \nonumber \\ 
	& \simeq & 4.19 \times 10^{9} \, {\rm year} \times \left( \frac{Q}{1.0} \right)^{-1} \left( \frac{\rho_{\rm par}}{3000 \, {\rm kg\,m^{-3}}} \right)  \left( \frac{r_{\rm par}}{1.0 \, {\rm m}} \right) \left( \frac{a_{\rm pla}}{1.0 \, {\rm au}} \right)^2 .
\end{eqnarray}
This expression is obtained directly from Eq.~\ref{eq_LH23_Eq20}. In the derivation we neglect the multiplicative factor enclosed in the large square brackets, \(\Bigl[\,\cdots\,\Bigr]\), in Eq.~\ref{eq_LH23_Eq20}. Over the allowed parameter ranges, this factor varies between \(\approx 0.46\) and \(1.5\) -- attaining the maximum for \(i\simeq0\), \(\epsilon\simeq0\), \(\phi=0\), and the minimum for \(i\simeq\pi/2\), \(\epsilon\simeq0\), \(\cos 2\phi=-3/4\). The square bracketed factor in Eq.~\ref{eq_LH23_Eq20} accounts for particle inclination $i$, planetary obliquity $\epsilon$, and the length of the planetary shadow $2\phi$. While omitting this bracketed factor changes the result only by an order-unity factor and is acceptable for the order-of-magnitude estimates intended here, including it yields a coefficient consistent with the earlier expressions of \citet{Bur79} and \citet{Gol82}, which did not consider planetary obliquity and planetary shadow.

Note that \(\tau_{\mathrm{decay}}\) is independent of the particle's initial circumplanetary distance $a_{\mathrm{par},0}$ when the term in the square brackets (Eq.~\ref{eq_LH23_Eq20}) has been approximated as unity. From Gauss's form of Lagrange's planetary equations, \(da/dt = 2F/n_{\mathrm{par}}\) for a tangential drag force \(F\). For a linear drag (\(F \propto v_{\mathrm{par}}\) with \(v_{\mathrm{par}} = n_{\mathrm{par}} a_{\mathrm{par}}\)), one obtains \(da_{\mathrm{par}}/dt \propto a_{\mathrm{par}}\). Consequently, particles that begin farther out (\(a_{\mathrm{par}}\) larger, \(n_{\mathrm{par}}\) smaller) experience a larger absolute semimajor-axis decay rate. Physically, outer orbits are more weakly bound, offer a longer torque lever arm, and spend more time per revolution under drag, all reinforcing the scaling \(da_{\mathrm{par}}/dt \propto a_{\mathrm{par}}\).

Figure~\ref{fig_2} maps the decay timescale \(\tau_{\mathrm{decay}}\) in the $a_{\mathrm{pla}}-r_{\mathrm{par}}$ parameter space. Particles of larger radius ($r_{\rm par}$ large) that orbit a planetary body farther from the Sun (\(a_{\mathrm{pla}}\) large) survive longest. The plot shows that gigayear-scale inward migration driven by Poynting-Robertson drag introduces a natural division in the solar system with regard to ring lifetimes: outer planetary bodies can retain rings, whereas inner planetary bodies lose them, consistent with the ringed-vs.-ringless pattern observed in the Solar System. We note that although the dimensionless radiation-pressure efficiency factor $Q$ sets the absolute scale of $\tau_{\mathrm{decay}}$, reasonable variations in $Q$ would not alter the global bimodal architecture of ringed versus ringless worlds.\\

\section{Discussion} 
\label{sec_discussion}

On the basis of the parameter space shown in Fig.~\ref{fig_2}, we may place new constraints on the formation and evolutionary history of the Solar System and predict the most promising locations for yet-undiscovered ringed worlds that future, higher-sensitivity observations may reveal. We emphasize that Fig.~\ref{fig_2} is intended as a general framework rather than being tied to any specific particle size. However, some observational constraints on ring particle sizes have been obtained for several planetary ring systems \citep[e.g.,][]{Bur01,Tis18}. For example, the typical sizes of particles in Saturn's dense rings are estimated to range from millimeters to meters \citep[e.g.,][]{Jer20}, and the particles in Uranus' dense rings are likely of comparable size \citep{Esp91}. In contrast, for fainter and more tenuous rings such as those around Jupiter and Neptune, the characteristic particle sizes are typically in the range of $1-100$ microns and these small particles are continuously replenished by ejecta from nearby moons generated by micrometeoroid impacts \citep[e.g.,][]{De18}. As future observations better constrain the typical particle sizes in different ring systems, or as readers adopt their own preferred values, our parameter-space maps can be directly applied to obtain the corresponding decay timescales and associated implications. Below, we provide some illustrative examples.\\

\subsection{Constraints on the Giant-Impact Origin of Satellite Formation}

Giant impacts are widely regarded as a common outcome of the final stages of planetary accretion \citep{Cha98,Agn99}. In the canonical scenarios proposed for the formation of Earth's Moon \citep{Can04} and Mars's moons \citep{Hyo17a}, such collisions generate a debris disk that subsequently spreads beyond the planet's Roche limit, allowing a satellite to accrete from the disk material \citep{Kok00,Ros16,Hes17}. Once the disks viscously spread and a moon system has formed, the residual disk becomes highly optically thin; individual particles then orbit in near-isolation and no longer experience the strong mutual interactions required for e.g., viscous spreading \citep{Cri25}. Alternatively, the ballistic transport \citep[e.g.,][]{Est23} may play an important role in the further dynamical evolution of optically thin disks, although it has not been proven that this process can completely remove the disk to be consistent with observations.

Giant impacts are thought to have occurred approximately $4 \, \mathrm{Gyr}$ ago, during the late stages of planetary accretion \citep[e.g.,][]{Qui16}. The proposed debris disk produced around Earth or Mars would have viscously spread in orders of magnitude shorter than the billions of years \citep{Kok00,Ros16}. An optically thin particulate ring --- composed of dynamically isolated particles --- could therefore, in principle, have persisted for gigayears, yet no such remnant is observed today.

Figure~\ref{fig_2} indicates that cumulative Poynting-Robertson drag can indeed remove the residual particle on gigayear timescales as long as the particle size is smaller than $\sim 0.1$~m, Therefore, to support the classical giant-impact origin of the Moon and the Martian satellites, future independent studies should investigate the characteristic size distribution of particles produced by giant-impact-generated debris disks during their subsequent spreading phase.

In contrast to the classical hypothesis, an alternative ring-satellite recycling scenario \citep{Hes17} suggests that Phobos is the latest survivor of several generations of satellites formed and disrupted only hundreds of millions of years ago assuming $\sim 0.1$~m sized particles, thereby leaving an optically thin ring of $\sim 0.1$~m particles that is likewise just hundreds of millions of years old. As Fig.~\ref{fig_2} shows, PR drag alone cannot clear $0.1$~m particles within such a short interval. To remain consistent with the absence of a present-day Martian ring \citep{Sho06}, that scenario therefore requires an additional removal mechanism, otherwise residual debris should still be detectable around Mars.

Venus and Mercury likely experienced giant impacts as well. Although the absence of satellites around these planets is beyond the scope of the present study\footnote[1]{\cite{Ato07} demonstrated that when the semi-major axis of a planet is small, the angular momentum of the satellite's orbit can be stripped away by solar tides, causing it to fall toward the planet.}, any resulting optically thin particulate disks would be even more susceptible to cumulative solar Poynting-Robertson drag, and could be cleared away on relatively short timescales even if the constituent particles were comparatively large (Fig.~\ref{fig_2}).\\

\subsection{Constraints on Rings Around Small Solar System Bodies}

Every solid planetary surface exhibits some impact craters, attesting to a history of collisions that inevitably eject material into orbit. Even so, neither the small asteroid Ryugu nor the much larger Vesta shows any evidence of a ring system, despite their well-documented crater populations. One possibility is simply that no single big event produced a sufficiently massive or long-lived debris disk to become optically detectable. Alternatively, a transient, optically thin ring by e.g. a small impact could have formed but was later removed by Poynting-Robertson drag. Whether such a ring would survive depends on both the epoch of its formation and the characteristic particle size: for example, particles generated early in Solar-System history, or composed of small particles, would be more easily cleared at Ryugu and Vesta locations (Fig.~\ref{fig_2}).

Radial distance from the Sun is a key parameter in Eq.~\ref{eq_decay}: a larger heliocentric semimajor axis $a_{\mathrm{pla}}$ yields a proportionally longer \(\tau_{\mathrm{decay}}\). Consequently, bodies in the outer Solar System are intrinsically more capable of retaining rings. Indeed, all giant planets have rings. Furthermore, this expectation is already borne out by recent detections of solid rings around several Centaurs \citep[e.g.,][]{Bra14,Ort15} and trans-Neptunian objects \citep[e.g.,][]{Ort17,Mor23,Per23}. We anticipate that an ever-increasing number of ring-bearing bodies will be discovered in the outer Solar System.

By the same argument, some Jupiter's Trojan asteroids may likewise host faint, as-yet-undetected rings. Because Trojan asteroids undergo mutual collisions on some timescales \citep{Mar22}, a relatively recent impact could have ejected debris that now persists as a transient particulate ring around the parent body. Based on the results shown in Fig.~\ref{fig_2}, for example, we can expect that icy/rocky particles of $\sim 1$\,cm in size can survive for billions of years, whereas particles of $\sim 1$\,mm in size can persist for about 100 million years. NASA's \emph{Lucy} mission will begin its tour of the Trojan swarms in 2027 and will provide a critical test of this prediction \citep{Lev21}.

Finally, ring particles can be generated at any epoch --- whether by impacts, tidal disruptions, rotational break-up, or other processes. Here an important point is that the size-frequency distribution typically has a negative slope, producing many small fragments and only a few large ones \citep[e.g.,][]{Doh69,Col96}. Smaller particles are removed on correspondingly shorter timescales: micron- to millimetre-sized grains are cleared not only by the Poynting-Robertson drag considered here but also by radiation pressure, electromagnetic forces, and collisional grinding \citep{Bur79}. Conversely, decimetre- to metre-scale and larger objects can survive in orbit for much longer intervals, yet their relative scarcity makes any associated ring system intrinsically difficult to detect. Therefore, considering the above arguments, observational biases --- particularly in the outer Solar System --- likely contribute to the underrepresentation of ringed bodies, and additional ring-bearing objects may yet be discovered.\\

\section{Conclusions and Future Perspectives}

We have mapped the solar Poynting-Robertson drag lifetime ($\tau_{\mathrm{decay}}$; Eq.~\ref{eq_decay}) of particles orbiting a central body --- whether a planet, dwarf planet, asteroid, or other small bodies --- as a function of the host body's heliocentric distance and the particle radius. The scaling \(\tau_{\mathrm{decay}}\propto a_{\mathrm{pla}}^{2} r_{\mathrm{par}}\) can naturally explain the Solar System's ringed-versus-ringless dichotomy: outer planets and several distant small bodies can retain rings for gigayears, whereas inner planets lose them on much shorter timescales. 

PR drag is only one of several non-gravitational forces, and particles smaller than those considered in this study are more susceptible to other perturbations. Additional perturbations --- such as collisional grinding, tidal or shepherding moon torques, electromagnetic forces, the asymmetric gravitational potential of the central body, and resonant interactions --- can further, or in some circumstances even predominantly, modify ring lifetimes across different systems. Furthermore, recycling between rings and moons can occur \citep{Hes19}, and the age of a ring's initial formation does not necessarily correspond to the ages inferred from other contexts -- such as its exposure age, structural age, dynamical age, or erosion age -- that we observe today \citep{Est23,Kem23,Hyo25}. The detailed discussion of these broader issues is beyond the scope of this paper; however, we refer interested readers to a recent review \citep{Cri25}.

We proposed that applying the PR drag framework may place quantitative constraints on planetary formation and evolution. For instance, any debris left by the impacts that produced the Moon or the Martian satellites may have consisted predominantly of sub-metre fragments in order to be removed within \(\sim4\;\mathrm{Gyr}\). Because \(\tau_{\mathrm{decay}}\) increases rapidly with heliocentric distance, we predict that additional ring-bearing objects await discovery in the outer Solar System, with the Jupiter-Trojan population providing an imminent observational test as NASA's \textit{Lucy} mission approaches its targets.\\

\begin{table*}[htbp]
    \centering
     \caption{Parameters used in this study. Here, the term “planet” refers to any celestial body that possesses a ring system.}
     \label{tab:parameters}
    \begin{tabular}{ll}
        \hline
        \textbf{Parameters} & \textbf{Meaning} \\
        \hline
        $R_{\mathrm{pla}}$ & radius of the planet \\[2pt]
        $a_{\mathrm{pla}}$ & semi-major axis of the planet around the Sun \\[2pt]
        $d$ & distance between the particle and the Sun \;($d\!\sim\!a_{\mathrm{pla}}$ is assumed) \\[2pt]
        $a_{\mathrm{par}}$ & semi-major axis of particle around the planet \\[2pt]
        $v_{\mathrm{par}}$ & particle velocity around the planet \\[2pt]
        $n_{\mathrm{par}}$ & particle mean motion around the planet \\[2pt]
        $r_{\mathrm{par}}$ & particle radius \\[2pt]
        $m_{\mathrm{par}}$ & particle mass \\[2pt]
        $\rho_{\mathrm{par}}$ & particle density \\[2pt]
        $i$ & particle inclination from planet's equatorial plane \\[2pt]
        $\phi$ & half the width of the planetary shadow\;($\phi\simeq\arcsin(R_{\mathrm{pla}}/a_{\mathrm{par}})$) \\[2pt]
        $Q$ & the dimensionless radiation-pressure efficiency factor \\[2pt]
        $F_{\mathrm{pla}}(a_{\mathrm{pla}})$ & insolation at planet's distance $a_{\mathrm{pla}}$ from the Sun \\[2pt]
        $\varepsilon$ & planet's obliquity \\[2pt]
        $c$ & speed of light \\[2pt]
        $L_{\odot}$ & solar luminosity\\
        \hline
    \end{tabular}
\label{table_1}
\end{table*}

\newpage
\bibliography{reference}{}
\bibliographystyle{aasjournalv7}



\end{document}